\documentclass{article}

\usepackage[comma,authoryear]{natbib}
\usepackage[utf8]{inputenc}

\usepackage{amsmath}
\usepackage{amssymb}
\usepackage{latexsym}
\usepackage{bm}
\usepackage[boxruled]{algorithm2e}
\usepackage{authblk}

\usepackage{ifpdf}

\ifpdf
\usepackage[pdftex]{graphicx}
\else
\usepackage{graphicx}
\fi

\graphicspath{{figures/}}

\date{Version of June 2, 2011}

\begin{document}

\newpage
\title{Modeling Non-Stationary Processes Through Dimension Expansion}

\renewcommand\Affilfont{\small} 
\author[1]{Luke Bornn}
\author[2]{Gavin Shaddick}
\author[1]{James V Zidek}
\affil[1]{Department of Statistics, University of British Columbia, Vancouver, Canada.} 
\affil[2]{Department of Mathematical Sciences, University of Bath, Bath, UK.}

\maketitle

{\bf Abstract}

In this paper, we propose a novel approach to modeling nonstationary spatial fields.  The proposed method works by expanding the geographic plane over which these processes evolve into higher dimensional spaces, transforming and clarifying complex patterns in the physical plane.  By combining aspects of multi-dimensional scaling, group lasso, and latent variable models, a dimensionally sparse projection is found in which the originally nonstationary field exhibits stationarity.  Following a comparison with existing methods in a simulated environment, dimension expansion is studied on a classic test-bed data set historically used to study nonstationary models.  Following this, we explore the use of dimension expansion in modeling air pollution in the United Kingdom, a process known to be strongly influenced by rural/urban effects, amongst others, which gives rise to a nonstationary field.

\medskip

\section{Introduction}\label{sect:1}
Recently there has been great interest in using spatial statistical methods to model environmental processes, with the aim of both gaining an improved understanding of underlying processes and making predictions at locations where measurements of a process are not available. The majority of such methods make the assumption that the underlying process is stationary (\cite{Cressie1993a}) which, for many   environmental systems, may be untenable.  

In this paper, we focus on accurately explicating the nonstationary structure that often arises in measurements of atmospheric, agricultural, and other environmental systems. If these systems share one underlying theme it is complex spatial structures, being influenced by such features as topography, weather, and other environmental factors. For example, the air quality characteristics of cities are likely to be more similar than that of rural areas irrespective of their geographic proximity. Ideally we might model these effects directly; however, information on the underlying causes is often not routinely available. Hence when modeling   environmental systems there exists a need for a class of models that are more complex  than those which rely on the assumption of stationarity.  

In the field of atmospheric science, empirical orthogonal functions have been used to model a nonstationary process as the sum of a stationary isotropic process and a set of basis functions with random coefficients representing departures from nonstationarity (\cite{Nychka1998a, Nychka2002a}).  Current approaches to modeling nonstationary processes in the statistical literature broadly comprise those that (i) combine locally stationary processes to create an overall nonstationary process and  (ii) `image warping'.  

A number of approaches for handling nonstationarity assume that over small enough spatial domains, the effects of nonstationarity are negligible, and hence locally stationary models may be used.  This concept is the basis of kernel approaches, early examples of which can be found in \cite{Haas1990a, Haas1990b}.   The process--convolution approach (\cite{Higdon1998a, Higdon1999a}) relies on the notion that a wide range of stationary Gaussian processes may be expressed as a kernel convolved with a Gaussian white noise process, with the kernel being allowed to vary spatially to account for nonstationarity.  The form of the kernel  allows for a broad expression of potential covariance functions, with a Gaussian kernel corresponding to a Gaussian covariance function and other choices of kernel resulting in other correlation structures.  Similarly, \cite{Fuentes2001a} suggested modeling the field  as a weighted average of local stationary processes within a set of regions, an idea which was later extended to include a continuous convolution of stationary processes (\cite{Fuentes2007a}). 
Various difficulties still remain in this class of models, including the lack of a complete and easily interpretable global model and the choice of local regions and details of the weight structures.  An alternative approach proposed by \cite{Sampson1992a} is that of ``image warping'', the central idea of which is that a nonstationary process may be stationary in a deformed, or warped, version of geographic space.  Multi-dimensional scaling (or related methods) can be used to find the deformed locations with a mapping between the original and deformed space found using, for example, a thin plate spline.


The principal idea underlying the proposed method is that of embedding the original field in a space of higher dimension where it can be more straightforwardly described and modelled.  Specifically, we shift the dimensionality of the problem from 2 or 3 dimensions to 4, 5, or more in order to recover stationarity in the process; we term our methodology ``dimension expansion.''  Our starting point is that nonstationary systems may be represented as low-dimensional projections of high-dimensional stationary systems (see, for example, \cite{Perrin2003a}). The method is superficially similar to that of image warping; however, it differs fundamentally in that here the locations in the geographic space are retained, with added flexibility obtained through the extra dimensions.  Additionally, it addresses one of the major issues with the image warping approach, namely folding of the space. This occurs in image warping if the  estimate of the function that transforms from geographical to deformed space is not injective.  As a result of folding, two geographically distinct locations may be mapped to the same location, meaning the variation between them will be incorrectly treated as measurement error and small scale variation (i.e. the nugget), which is expressly appropriate only for collocated and other proximal monitoring sites.  In such cases, mapping quantities such as prediction intervals becomes particularly challenging both in terms of implementation and interpretation.

The remainder of the paper is organised as follows: Section 2 introduces the dimension expansion framework proposed here, including an illustrative example to demonstrate the fundamental concepts behind the approach.  This example is then used to draw comparisons to image warping.  In Section 3, dimension expansion is applied to two real life examples.  First, the solar radiation dataset originally used in \cite{Sampson1992a} and used as a test-bed in various more recent image warping papers is studied.  Second, we study air pollution from seventy-seven monitoring locations in the United Kingdom which show clear signs of nonstationarity.  We highlight the ability of dimension expansion to accurately model such data as measured through cross-validated prediction error. Finally, Section 4 provides a discussion and suggestions for future developments.

\section{Dimension Expansion}\label{sect:2}

While early work (\cite{Cressie1993a}) dealt primarily with stationary models, it is now generally recognized that many spatial processes $\{Y(\bm{x}) : \bm{x} \in \mathcal{S}\} ,\hspace{2mm} (\mathcal{S} \in \mathcal{R}^d)$ fail to satisfy this assumption.  Environmental systems might exhibit behaviour that looks locally stationary, yet when considered over large and heterogenous domains they very often exhibit nonstationarity.  For ease of notation, we consider $Y(\bm{x})$ to be a (potentially nonstationary) mean-zero Gaussian process and place our emphasis on representing the nonstationary structure.

A principal task in spatial statistics is estimating a variogram model (or correlation function) to explain spatial dependencies. The theoretical variogram, defined as 
\begin{align*}
	2 \gamma (\bm{x_i},\bm{x_j}) = E \left( | Y(\bm{x_i}) - Y(\bm{x_j}) |^2 \right)
\end{align*}
is typically modeled using a parametric stationary variogram $\gamma_\phi(\bm{h})$ depending only on $\bm{h} = \bm{x_i} - \bm{x_j}$, the difference vector between locations, and the parameter(s) $\phi$.  If the field is nonstationary, such a model will be a misspecification.  In response, we transform the set of observed spatial locations $\mathcal{S} \in \mathcal{R}^d$ into one of higher dimension $\mathcal{S'} \in \mathcal{R}^{d+p}$, where $p>0$ and $\mathcal{S}$ is a subset of the dimensions of $\mathcal{S'}$.  Put plainly, such an approach amounts to allowing extra dimensions for the observed locations $\bm{x_1}, \dots, \bm{x_s}$, notated as $\bm{z_1}, \dots, \bm{z_s}$ such that the field $Y([\bm{x},\bm{z}])$ is stationary with a variogram model  $\gamma_\phi([\bm{x_i},\bm{z_i}] - [\bm{x_j},\bm{z_j}])$.  Here $[\bm{x},\bm{z}]$ is the concatenation of the dimensions $\bm{x}$ and $\bm{z}$.

\citet{Perrin2003a} explore this idea in the particular case where both the covariance function and the expansion from $\bm{x}$ to [$\bm{x}$, $\bm{z}$] are known.  In their motivating example, they consider the following stationary covariance on the plane:
\begin{align*}
	cov(Y([\bm{x_i},\bm{z_i}]), Y([\bm{x_j},\bm{z_j}])) = \exp(-|\bm{x_i} - \bm{x_j}| - |\bm{z_i} - \bm{z_j}|).
\end{align*}
By restricting to the set $\bm{z} = \bm{x}^2$ and defining $Y'(\bm{x}) = Y([\bm{x},\bm{x}^2])$, the resulting covariance function on this reduced-dimension field is nonstationary, namely
\begin{align*}
	cov(Y'(\bm{x_i}), Y'(\bm{x_j})) = \exp(-|\bm{x_i} - \bm{x_j}|) \exp(1 + |\bm{x_i} + \bm{x_j}|).
\end{align*}
\citet{Perrin2003a} then consider the reverse problem, proving that a nonstationary random field indexed by $\mathcal{R}^d$ (with moments of order greater than 2) can always be represented as second-order stationary in $\mathcal{R}^{2d}$.  It is not, however, necessary to move from $\mathcal{R}^d$ to $\mathcal{R}^{2d}$ to obtain the existence of a stationary field.  Consider a recent result of \citet{Perrin2007a}, which states that a Gaussian random vector can always be interpreted as a realization of a stationary field in $\mathcal{R}^p, p \ge 2$, subject to moment constraints on the vector.  From this it is straightforward to state that, similarly, a realization of a Gaussian process in $\mathcal{R}^d$ may be interpreted as a realization of a stationary field in $\mathcal{R}^{d+p}, p \ge 2$ (similarly, subject to moment constraints), with the covariance function ignoring the initial $d$ dimensions. 

The above results show the existence of higher-dimensional stationary representations for nonstationary fields, yet in the vast majority of situations neither a nonstationary variogram, nor an analytic mapping to higher dimensions, is known. Here we construct a framework for using higher-dimensional representations to model nonstationary systems, with the goal of learning the latent dimensions nonparametrically from information contained within the data.

To learn the expanded, or latent, dimensions $\bm{z_1}, \dots, \bm{z_s}$ we propose
\begin{align}
\hat{\phi}, \bm{Z} = \underset{\phi,\bm{Z'}}{\operatorname{argmin}} \sum_{i<j}(v^*_{i,j} -  \gamma_\phi(d_{i,j}(\left[\bm{X},\bm{Z'}\right])))^2
\label{eq:optim_vario}
\end{align}
where $v^*_{ij}$ estimates the spatial dispersion between sites $i$ and $j$, for example
\begin{align*}
	v^*_{ij} = \frac{1}{|\tau|}\sum_{\tau} |Y(\bm{x}_i) - Y(\bm{x}_j)|^2,
\end{align*}
with $\tau>1$ indexing multiple observations of the system, the handling of which is  considered in the discussion, and $d_{i,j}(\left[\bm{X},\bm{Z}\right])$ is the $i,j^{th}$ element of the distance matrix of the (augmented) locations $[\bm{X},\bm{Z}]$.  Once the matrix $\bm{Z} \in \mathcal{R}^{s} \times \mathcal{R}^p$ is found, a function $f$ is built such that $f(\bm{X}) \approx \bm{Z}$.  While a wide range of options exist, we follow \citet{Sampson1992a} in using thin plate splines, here one for each dimension of $\bm{Z}$.  The smoothing parameter of the thin plate spline (denoted $\lambda_2$) is used to control the smoothness of the resulting warped space through penalization of the bending energy
\begin{align*}
	\int_{\mathcal{R}^2} \left[  \left( \frac{\partial^2 f}{\partial^2 x_1} \right)^2 +  2\left( \frac{\partial^2 f}{\partial x_1 \partial x_2} \right)^2 + \left( \frac{\partial^2 f}{\partial^2 x_2} \right)^2     \right]  d x_1 d x_2, \mbox{   (for $d=2$)},
\end{align*}
and is therefore analogous to $\lambda_{IW}$, the thin plate spline parameter in the image warping framework.  Setting $\lambda_2 = 0$ results in an interpolating spline, whereas $\lambda_2 \rightarrow \infty$ results in the linear least squares fit.  The nonlinear functions $f$ are therefore linear combinations of basis functions centered at the locations $\mathcal{S} \in \mathcal{R}^d$.  Once a model is built in the expanded space, $f^{-1}$ will bring us from the manifold in $\mathcal{R}^{d+p}$ defined by $(\bm{X},f(\bm{X})), \bm{X} \in \mathcal{R}^d$ back to the original space.  

Due to our unique formulation, we have $f^{-1}(\bm{Z}) = \bm{X}$, and we need not be concerned with the difficulties associated with ensuring that $f$ is bijective as in earlier approaches.  Thus we may view the originally observed locations $\bm{X}$ as a projection from a manifold within a higher dimensional space, $\left[\bm{X}, \bm{Z}\right]$, in which the process is stationary.  As an obvious (and direct) example, a process which is stationary given both geographical location and elevation may result in a nonstationary field given only longitude and latitude.  Learning the latent dimensions (whether or not they have a physical meaning, such as elevation) means that a stationary model may be used in the expanded space.

In many situations, it is unclear how many additional dimensions are needed to accurately model the spatial field.  One could use cross-validation or a model selection technique to determine the dimensionality of $\bm{Z}$; however, recognizing that (\ref{eq:optim_vario}) might result in overfitting the spatial dispersions, we would also like to regularize the estimation of $\bm{Z}$.  As a result, we modify (\ref{eq:optim_vario}) by including a group lasso penalty term on $\bm{Z}$, where the groups are the dimensions of $\bm{Z}$ (\cite{Yuan2006a}).  The resulting objective function is
\begin{align}
\hat{\phi}, \bm{Z} = \underset{\phi,\bm{Z'}}{\operatorname{argmin}} \sum_{i<j}(v^*_{i,j} -  \gamma_\phi(d_{i,j}(\left[\bm{X},\bm{Z'}\right])))^2  + \lambda_1 \sum_{k=1}^{p} ||\bm{Z'}_{\cdot,k}||_1.
\label{eq:optim_vario_L1}
\end{align}
where $\bm{Z'}_{\cdot,k}$ is the $k$-th column (dimension) of $\bm{Z'}$.  As a consequence of this revised objection function, one need only determine a maximum number of dimensions $p$ and the parameter $\lambda_1$, whereupon the learned augmented dimensions $\bm{Z}$ will be both regularized towards zero and sparse in dimension.  Hence $\lambda_1$ can be viewed as regularizing the estimation of $\bm{Z}$ and determining the dimension of the problem, whereas $\lambda_2$ controls the smoothness of the augmented dimensions; we suggest learning both through cross-validation, although other model fit diagnostics or prior information may be used as well.

It is relatively straightforward to solve (\ref{eq:optim_vario_L1}) using the gradient projection method of \citet{Kim2006a}, which conducts block-wise updates for group lasso with general loss functions.  Here the blocks are the dimensions of $\bm{Z}$, and hence the optimization is efficient even for a large number of spatial locations.  Optimization details are given in the appendix.  For ease of exposition we use an exponential variogram,
\begin{align*}
\gamma_\phi (\bm{x_1}, \bm{x_2}) = \phi_1 (1- \exp(-||\bm{x_1} - \bm{x_2}||/\phi_2)) + \phi_3,
\end{align*}
which works well in the examples that follow, although the method applies analogously to other variograms.

\subsection{Illustrative example}

\begin{figure}
\centering
\includegraphics[width=0.95\textwidth]{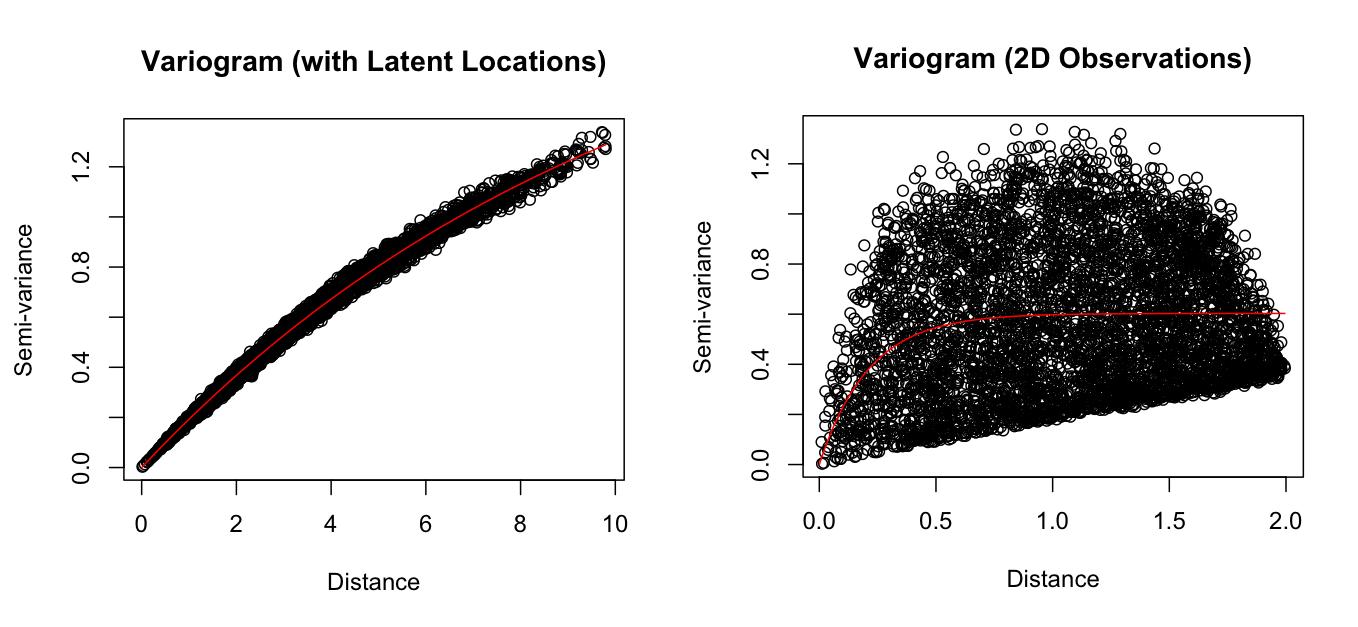}
\caption{Empirical variograms from the original process (left) as well as a 2-D projection (right) on the illustrative ellipsoid example.  A fitted exponential variogram is shown by the solid line.}
\label{Sim_Varios}
\end{figure}

\begin{figure}
\centering
\includegraphics[width=0.95\textwidth]{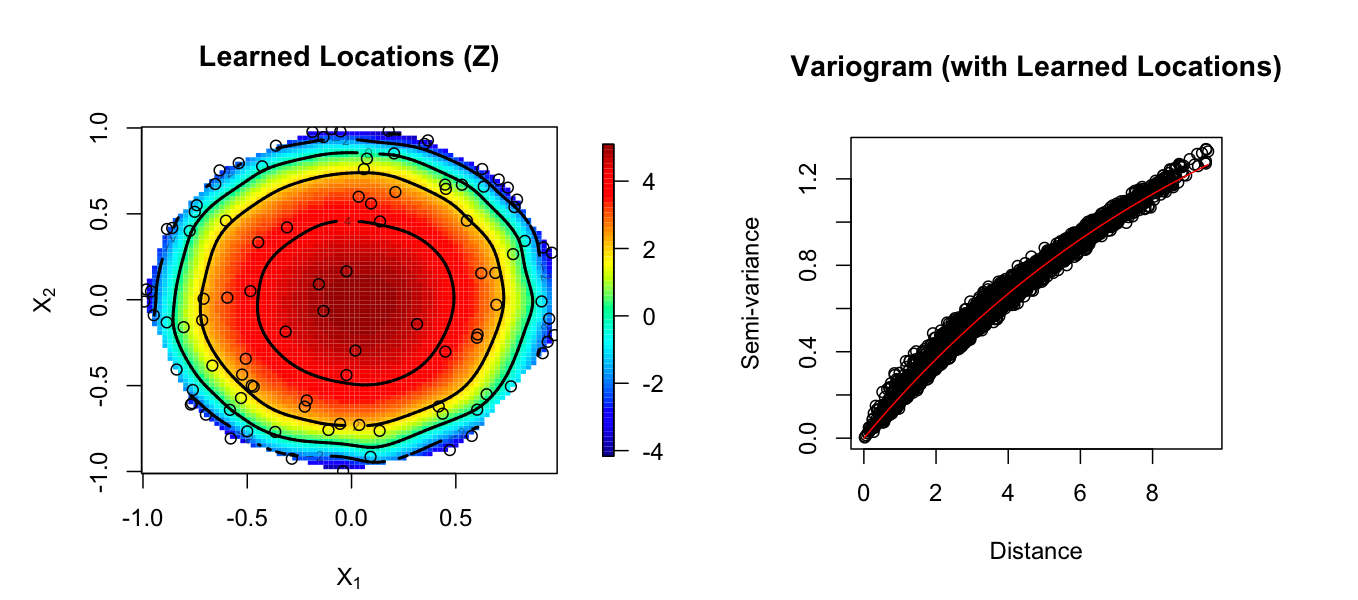}
\caption{Learned latent locations (left) using $\lambda_2 = 10^{-4}$ as well as corresponding empirical variogram (right) after dimension expansion is applied.  The fitted exponential variogram is shown by the solid line.}
\label{Sim_Learned}
\end{figure}

We now present an illustrative example to help explain the concepts behind this proposed dimension expansion approach, as well as demonstrate the inability of image warping to handle complex nonstationarity.  Specifically, we simulate a Gaussian process with $s=100$ locations on a $3$-dimensional ellipsoid centered at $(0,0,0)$ such that the projection to the first $2$ dimensions is a disk centered at the origin.  Figure \ref{Sim_Varios} plots the empirical variograms for the original $3$-dimensional space as well as the $2$-dimensional projection, the latter of which results in a highly noisy empirical variogram cloud.
Our goal is to recover the lost dimension through dimension expansion by optimizing (\ref{eq:optim_vario_L1}) with $\lambda_1=0.5$, chosen to induce $\bm{Z}$ to have one dimension.  Here we calculate the matrix of empirical dispersions $v^*_{ij}$ using $1000$ realizations of the Gaussian process.  Figure \ref{Sim_Learned} shows the resulting learned locations as well as the corresponding empirical variogram, where we see that dimension expansion is capable of recovering the lost dimension, resulting in a variogram that closely reproduces the original.

\subsection{Image Warping and Folding}

In the image warping approach, \citet{Sampson1992a} employ non-metric multidimensional scaling to move the locations along the geographic space, followed by fitting of the variogram $\gamma_\phi$ using traditional variogram fitting methods.  From this, a function $f$ is found to go from the original to the warped locations, and back via $f^{-1}$.  A number of adaptations of this approach have been proposed. \cite{Smith1996a} proposed modeling the covariance function as a linear combination of radial basis functions  using maximum likelihood (as suggested by \cite{Mardia1993a}).  \cite{Monestiez1991a} and \cite{Monestiez1993a} noted that the multi-stage algorithm of Sampson and Guttorp does not correspond to a unified optimization problem and  instead propose finding the locations and fitting the variogram using a single objective function, an approach also pursued by \cite{Meiring1997a}.  It is worth noting that \cite{Monestiez1991a} also explore mappings from $\mathcal{R}^2$ to $\mathcal{R}^3$ in the context of analyzing acid rain data, as the same-dimension mapping was incapable of describing the nonstationarity arising in the observed field.  In a similar vein, \cite{Iovleff2004a} propose using simulated annealing to fit the spatial deformation model.  Rather than imposing smoothness on the deformation through thin plate splines, they use Delauney triangulation to constrain the mapping $f$ from folding on itself.  In order to acknowledge  the uncertainty associated with the deformed locations, \cite{Damian2001a}, \cite{Schmidt2003a}, and recently \cite{Schmidt2011a} have proposed Bayesian implementations of this approach, the latter additionally using observed covariate information to warp into higher dimensions. 

\begin{figure}
\centering
\includegraphics[width=0.95\textwidth]{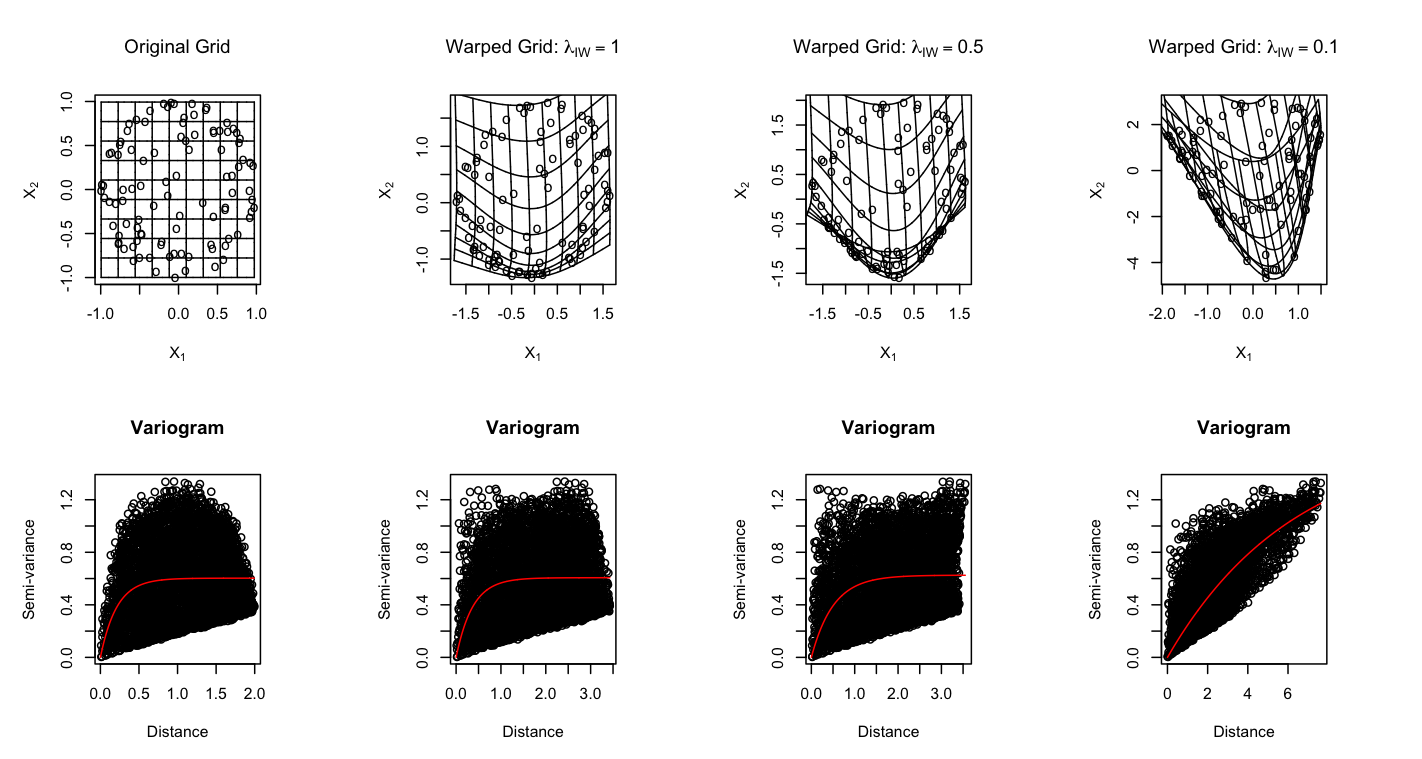}
\caption{Warped grid of locations (top) and corresponding variograms (bottom) for various settings of the thin plate spline parameter $\lambda_{IW}$ using the image warping technique of \citet{Sampson1992a}.}
\label{Sim_SG}
\end{figure}

As described in the introduction, the image warping framework can suffer from problems of folding, namely of $f$ not being bijective (See \citet{Zidek2000a} for a particularly extreme example of folding).  Considering the illustrative example of Section 2.1, admittedly designed to be illustrative of such folding, we apply the image warping technique (\cite{Sampson1992a}) with $f$ modeled as a thin plate spline.  Because the image warping framework contains no term similar to $\lambda_1$ to regularize the warped locations, smoothing must be done through the thin plate spline parameter $\lambda_{IW}$ (analogous to $\lambda_2$ in the proposed dimension expansion framework).  Figure \ref{Sim_SG} shows the warped grids and resulting empirical variograms for various settings of $\lambda_{IW}$ applied to the ellipsoid example introduced in Section 2.1.  We observe immediately that for a highly penalized $f$ (corresponding to large $\lambda_{IW}$) the space does not fold; however, the variogram fit is very poor.  As $\lambda_{IW}$ is relaxed to improve the fit, the space begins to fold, highlighting a potentially serious problem with the image warping framework -- an issue which is addressed in the dimension expansion paradigm proposed here.

Also related to the proposed dimension expansion method are latent space models such as that proposed by \citet{Hoff2002a}.  Here, latent dimensions are used to help learn a network of relationships between individuals.  Recent work in the field of spatial statistics has also exploited latent dimensions to ensure valid cross-covariance functions in multivariate fields.  Specifically, \citet{Apanasovich2010a} use latent dimensions for the different variables in order to build a class of valid cross-covariance functions.

\section{Applications}\label{sect:3}

We now present two applications of dimension expansion applied to the modeling of nonstationary processes using two real data sets. The first uses  the solar radiation data (\cite{Hay1984a}) studied in the original image warping paper of \citet{Sampson1992a}. The second consists of measurements from a network of air pollution (black smoke) monitoring sites in the UK, further details of which can be found in \cite{Elliott2007a}.

\subsection{Solar radiation}
\begin{figure}
\centering
\includegraphics[width=0.95\textwidth]{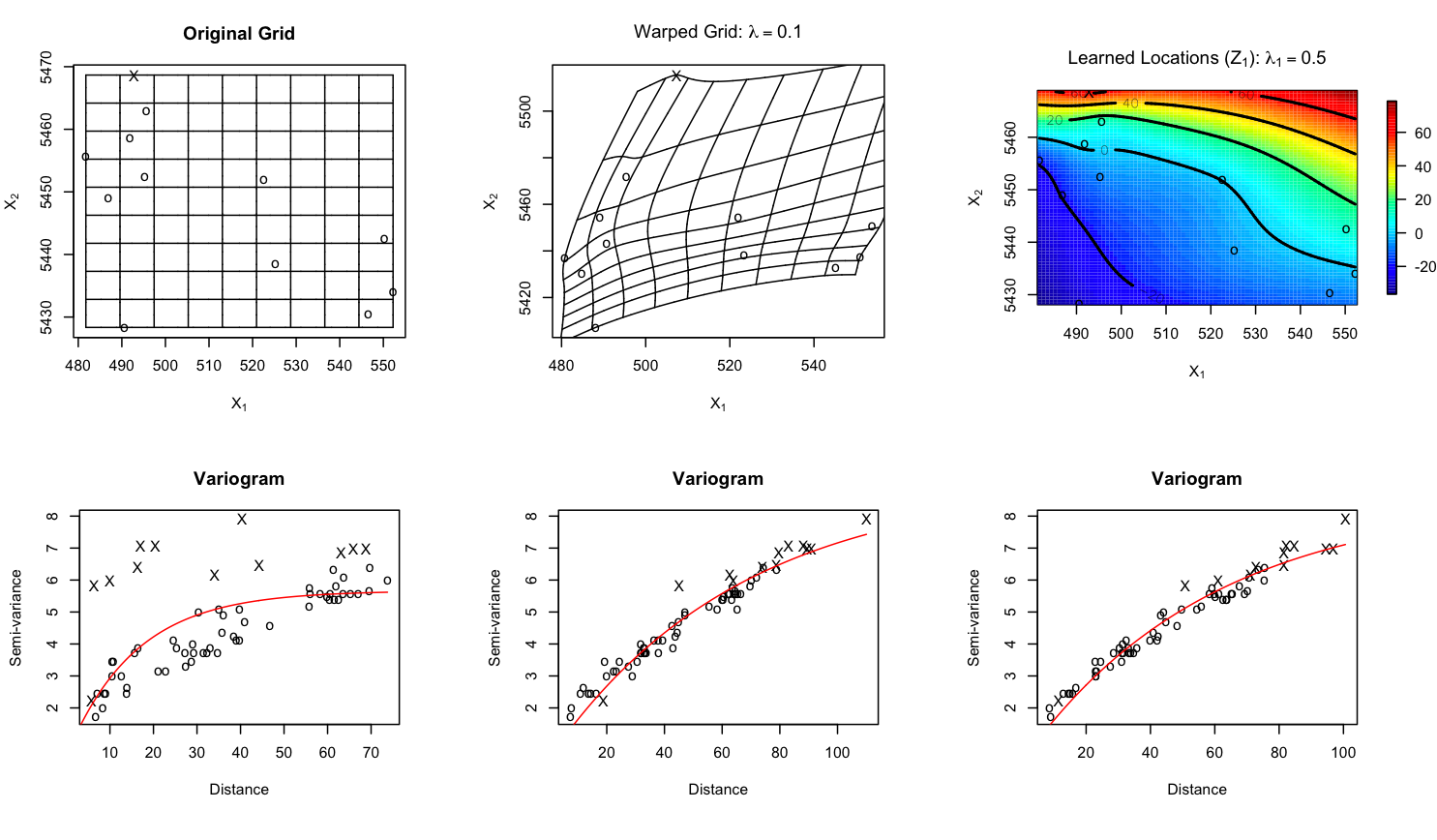}
\caption{Original locations and empirical variogram for the solar radiation data (left); warped locations and associated empirical variogram using image warping with $\lambda_{IW}=0.1$ (centre); learned locations with associated empirical variogram using dimension expansion with $\lambda_1=0.5, \lambda_2=10^{-4}$ (right).  The units for the semi-variance are $(M J m^{-2} \mbox{day}^{-1})^{2}$, and for distance are $km$ (UTM coordinates, divided by $1000$).  The fitted variogram is shown by a solid line, and points associated with Grouse mountain (station 1) are highlighted using an ``X''.}
\label{Solar_SG_and_DW}
\end{figure}

   The data of \citet{Hay1984a} includes measurements of solar radiation from $12$ stations in the area surrounding Vancouver.  Due to the location and elevation of station $1$ (Grouse mountain), the field is inherently nonstationary, as exhibited by the sample variogram (Figure \ref{Solar_SG_and_DW}).  This figure shows the original and warped locations using Sampson and Guttorp's image warping approach with corresponding variogram plot.  Image warping moves the locations (in particular the station at Grouse mountain, which is the northernmost location) to achieve something closer to stationarity.  It is worth noting that overfitting may be controlled through the parameter $\lambda_{IW}$ of the thin plate spline.

\begin{figure}
\centering
\includegraphics[width=0.95\textwidth]{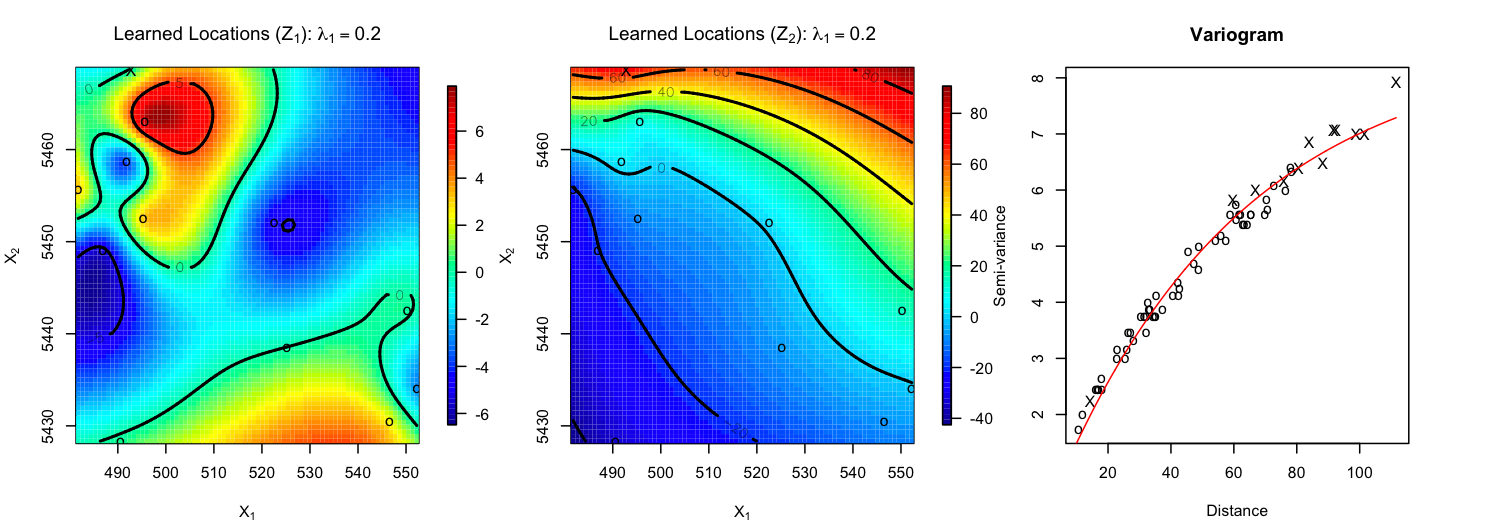}
\caption{Dimension expansion of the solar radiation surface using $\lambda_1=0.2, \lambda_2=10^{-4}$.  $Z$ here is 5 dimensions, with $Z_3,Z_4$, and $Z_5$ set to zero as a result of the sparsity-inducing penalization.  The first two panels show the learned locations, and the right-most panel shows the associated empirical variogram (fitted variogram shown in red).  The units for the semi-variance are $(M J m^{-2} \mbox{day}^{-1})^{2}$, and for distance are $km$.  Points associated with Grouse mountain (station 1) highlighted using an ``X''.}
\label{Solar_4D}
\end{figure}

Figures \ref{Solar_SG_and_DW} and \ref{Solar_4D} show the results of applying the dimension expansion approach using $\lambda_1=0.5$ and $\lambda_1=0.2$, respectively, using a maximum number of dimensions of $p=5$.  The original locations are shown, as well as the added dimensions ($\bm{Z}$).  With $\lambda_1=0.5$ (Figure \ref{Solar_SG_and_DW}, right), dimension expansion adds one additional dimension which primarily serves to push Grouse mountain out of the plane, reflecting the a priori suggestion that it is elevation that leads to the station's spurious correlation pattern.  Interestingly, the contours of the learned dimension closely resemble the elevation contours of the mountains surrounding the Vancouver area.  With $\lambda_1=0.2$ (Figure \ref{Solar_4D}), $2$ extra dimensions are used, and the fit of the parametric variogram improves marginally.  We can therefore see how $\lambda_1$ influences the number of extra dimensions used, as well as the shrinkage in each dimension, in order to control the level of fit.

\subsection{Air pollution}
\begin{figure}
\centering
\includegraphics[width=0.95\textwidth]{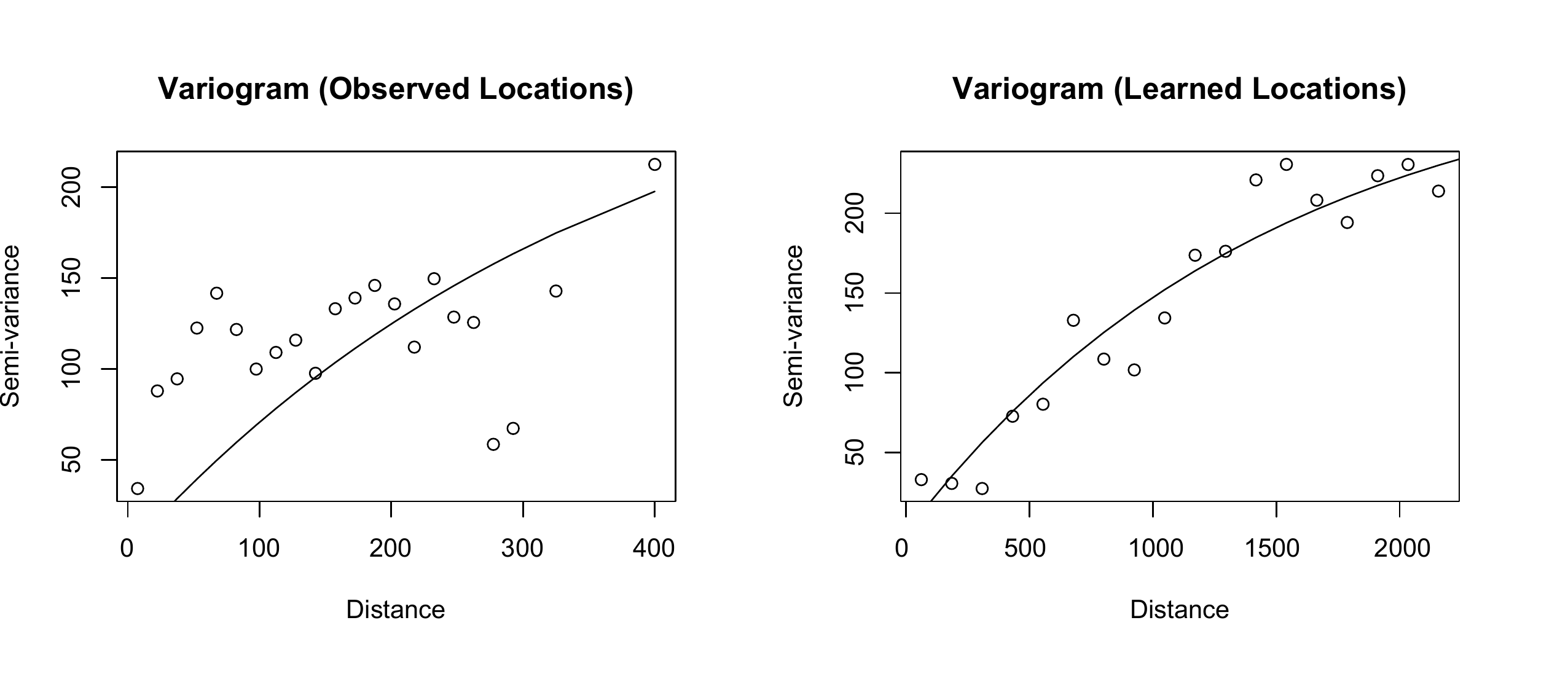}
\caption{Binned empirical and fitted (solid line) variograms on the UK black smoke data, following dimension expansion.  In the original geographic space, we see a dip in the empirical variogram at roughly $280 km$, corresponding to the distance between London and Liverpool/Manchester.  After dimension expansion is applied, the ridge between London and Liverpool/Manchester removes this effect of nonstationarity.  The units for the semi-variance are $(\mu$gm$^{-3})^2$ and for distance are $km$.}
\label{UK_Vario}
\end{figure}

The data consists of annual average concentrations of black smoke ($\mu$gm$^{-3}$) over a period of sixteen years from 77 locations within the UK operating between April 1978 and March 1993 (inclusive) and was obtained from the Great Britain air quality archive (www.airquality.co.uk). Sites were selected in areas defined wholly or partially residential and measurements were aggregated to ward level (based on the 1991 census) using a geographical information system (\cite{Elliott2007a}). The majority of wards contained a single site, but where there were more than one, records were either joined together if the time periods did not overlap or averaged if time periods of operation were simultaneous.  Due to similarities in levels of air pollution in urban locations, even if they are not geographically close, the field is known to be nonstationary.  Specifically, we see in Figure \ref{UK_Vario} reduced empirical dispersions for distances around $280-290 km$ (the distance between London and Liverpool/Manchester), indicating that these urban centers are more similar than their distances would suggest.  Our goal is to uncover and explore this nonstationarity through the dimension expansion framework.

\begin{figure}
\centering
\includegraphics[width=0.5\textwidth]{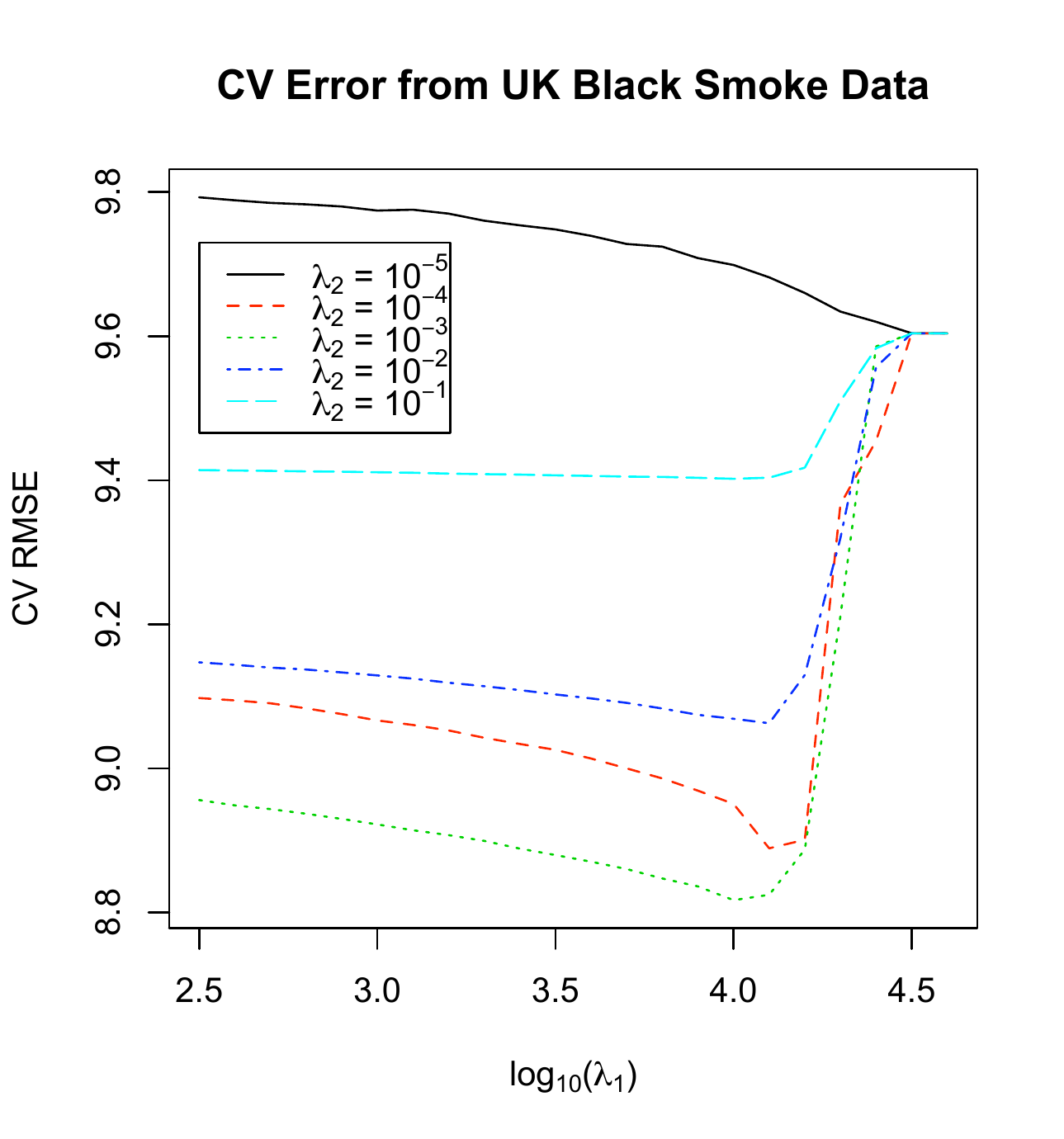}
\caption{Leave-10-out cross-validated prediction error  of dimension warping applied to the UK black smoke data.  Here we see optimal prediction for moderate values of both $\lambda_1$ and $\lambda_2$.}
\label{UK_CV}
\end{figure}

We begin with cross-validation to learn the optimal setting of the parameters $\lambda_1, \lambda_2$ using (\ref{eq:optim_vario_L1}) as described in Section 2.  Figure \ref{UK_CV} shows the resulting cross-validation RMSE for various parameter settings.  We can see that moderate values of both $\lambda_1$ and $\lambda_2$ result in the best prediction performance.
As $\lambda_1$ increases to its highest value ($10^{4.5}$), no dimension expansion occurs, and hence $\lambda_2$ has no impact.  From this it is straightforward to see that the use of the original geographic space is a special case of the dimension expansion framework.

\begin{figure}
\centering
\includegraphics[width=0.5\textwidth]{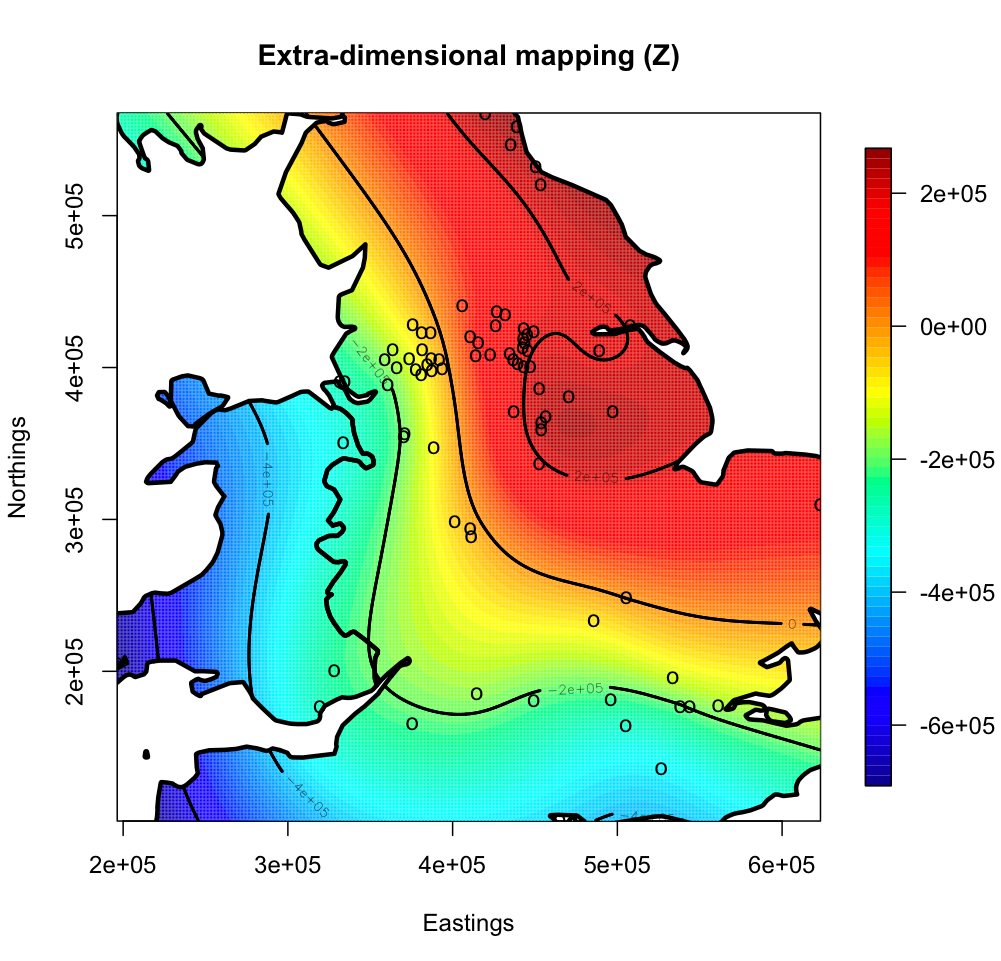}
\caption{Map of the learned dimension following dimension expansion, which has found a strong ridge connecting the major cities, indicating closer correlation between these locations than would be suggested in geographic space.}
\label{UK_Map}
\end{figure}

Using these parameter values, the dimensionally sparse optimization (\ref{eq:optim_vario_L1}) used by dimension expansion leaves all but one dimension of $\bm{Z}$ set to zero.  This dimension is shown in Figure \ref{UK_Map}, where we see a strong ridge connecting London, Birmingham, Liverpool, and Manchester.  Hence in the extra dimension major cities are moved closer together while rural areas are pushed further away.
The variograms before and after the dimension expansion are shown in Figure \ref{UK_Vario}, where we see no indications of nonstationarity after dimension expansion is applied.

\section{Discussion}

By augmenting the dimensionality of the underlying geographic space, we have developed a highly flexible approach for handling the nonstationarity that often arises in environmental systems.  While ostensibly similar to image warping, the proposed method avoids the issue of folding and allows one to model much more complex nonstationarity patterns through interdimensional expansions, allowing the user to perform nonparametric learning of the mapping function.  In addition, through the use of a group lasso penalty, we are able to estimate the number of augmented dimensions, as well as regularize the optimization problem.  Lastly, we have highlighted the practical application of the dimension expansion approach in three examples, two of which use data from observed environmental processes.  It is worth noting that while we have developed the spatial model in terms of variograms, it could alternatively be expressed in terms of covariances; see, for example, \citet{Gneiting2001a} for a thorough comparison of the two approaches.

In general, models will comprise a spatial mean or ‘trend’ term together with spatial covariance for deviations from this trend. It is desirable to maximally reflect the variation in the response using the mean function and thus known covariates, but inevitably the mean function will not be able to capture all of the spatial variation and thus residual spatial variation must be modeled specifically. When all relevant covariates are included in the mean term, it is commonly assumed that the resulting spatial term is stationary. However, as the Karhunen-Loeve expansion shows, the modeling of spatial trend and covariance are inseparable and misspecification of the former will induce a second order distortion in the latter, thus violating any assumptions of stationarity in many cases.  Due to the complexity of environmentally processes, mis-specification is inevitable because all relevant covariates can never be known or, even if known, observed. In the air pollution example presented here, concentrations in cities appeared to be more similar than that of rural areas irrespective of their geographic proximity. If available, it would be possible to incorporate a measure of rurality in the mean function, possibly produced using a geographical information system based on population density data. However, even if such information were available, stationarity would still not be guaranteed and so there is a need for methods such as that proposed here to allow nonstationary models to be constructed for the spatial process.

A Bayesian image warping approach which allows covariate effects to be included in the correlation structure has recently been suggested by \citet{Schmidt2011a}.  By treating covariates as analogous to geographic coordinates, they warp the combined location-covariate space into a deformed space of the same dimension.  To achieve computational efficiency, they consider a special case which restricts the form of the possible mapping function and assumes the spatial process to be a 2D manifold with covariates treated as separate values at each location.

In practice, environmental data will often take the form of a number of measurements made over time at each location rather than true spatial replications per se. In order to try and isolate the purely spatial part of the process, the mean function may incorporate a temporal component into the mean function, modelling underlying temporal patterns and allowing the possibility of time-varying covariates, or even space-time interactions. In the absence of such covariate information, it would be possible to consider the notion of time-varying nonstationarity structure.  For instance, if one wants to study the changing impact of cities and industrial areas on air pollution levels, examining changes in stationarity over time would be a valuable way to understand these changes. The dimension expansion framework is also amenable to multivariate extensions.  We are currently exploring a scenario whereby the dimension expansion functions and locations have a hierarchical structure, allowing the dimension expansion to vary for different variables, yet be tied together through the hierarchy.

We have demonstrated how the proposed approach can be used to perform predictions in the transformed, stationary space and mapped back to the original space. At present the choice of the mapping, learning of latent locations, and prediction are performed in isolation. As the Sampson and Guttorp (1992) approach was set within a Bayesian framework by \citet{Damian2001a} and \citet{Schmidt2003a}, setting the current algorithmic approach within such an inferential framework would allow the inherent uncertainty to be accurately reflected in resulting inferences and this is the goal of current research.

\bibliographystyle{chicago}
\bibliography{lukebornn}

\appendix

\section{Optimization of Equation (\ref{eq:optim_vario_L1})}

As with traditional multi-dimensional scaling, penalization functions of the form (\ref{eq:optim_vario}) do not have a unique maximum.  However, the learned locations are unique up to rotation, scaling, and sign.  The optimization problem (\ref{eq:optim_vario_L1}) is more regularized, however, due to the presence of the $l_1$-norm.  Specifically, not all rotations and scalings of the learned locations will have the same $l_1$-norm, and hence the resulting optimization is unique only up to sign and indices of zero/non-zero dimensions.

In our experience, traditional optimization procedures such as Nelder-Mead or the Broyden–Fletcher–Goldfarb–Shanno method (\cite{Nocedal1999a}) work well for a moderate amount of locations ($s<100$) and dimensions ($p<3$).  For larger problems, it may be necessary to use purpose-built optimization routines intended for generalized group lasso.  Let $\Omega(\bm{U})$ be the first term in equation \ref{eq:optim_vario_L1}, where $\bm{U} = [\bm{X},\bm{Z}]$.  Then column $k$ of the gradient matrix is
\begin{align*}
	\nabla_k {\Omega(\bm{U})} = \frac{2}{p} \bm{\Gamma} \circ \left( \bm{U}_{\cdot,k} \bm{1}_{p \times p} - \bm{1}_{p \times p} \bm{U}_{\cdot,k}^T \right) \bm{1}_{p \times 1}
\end{align*}
where
\begin{align*}
	\bm{\Gamma}_{i,j} = \left( \gamma_\phi(d_{ij}(\bm{U})) - \nu_{ij}^* \right)  \frac{\partial \gamma_\phi}{\partial d_{ij}}.
\end{align*}
Using this gradient information, the gradient projection algorithm of \citet{Kim2006a} may be used to optimize (\ref{eq:optim_vario_L1}).  The algorithm proceeds as follows:

\begin{algorithm}[H]
	Initialize : $\bm{U}^0= \bm{0}$, $\alpha$ : sufficiently small positive constant \\
\For{$t=1,\dots,T$}{
	Set $\bm{u} = \bm{U}^{t-1} - \alpha \nabla {\Omega(\bm{U}^{t-1})}$ and $\eta = \{ 1,\dots,p \}$\\
	\While{$M_j > 0 \hspace{3mm}\forall j$}{
		For $j=1,\dots,p$
		\begin{align*}
			M_j = I(j \in \eta) \times \left( ||\bm{u}_j|| + \frac{M - \sum_{j \in \eta} ||\bm{u}_j||}{|\eta|}    \right)
		\end{align*}
		Set $\eta = \{ j: M_j > 0 \}$
	}
	Set $\bm{U}_{\cdot,j}^{t-1} = \bm{u}_j \frac{M_j}{||\bm{u}_j||}$ for $j= 1,\dots,p $
}
Return $\bm{U}^{T}$
\end{algorithm}
Further algorithmic details, such as the tuning of $M$ and the setting of the algorithmic parameter $\alpha$, can be found in \citet{Kim2006a}.

\end{document}